\documentclass[fleqn,twoside]{article}
\usepackage{espcrc2}

\usepackage{graphicx}
\usepackage[figuresright]{rotating}

\newcommand{\AmS}{{\protect\the\textfont2
  A\kern-.1667em\lower.5ex\hbox{M}\kern-.125emS}}
\newcommand{\beqn}{\begin{eqnarray}}
\newcommand{\eeqn}{\end{eqnarray}}
\newcommand{\be}{\begin{equation}}
\newcommand{\ee}{\end{equation}}

\def\beq{\begin{equation}}
\def\be{\begin{equation}}
\def\beqn{\begin{eqnarray}}
\def\ee{\end{equation}}
\def\eeq{\end{equation}}
\def\eeqn{\end{eqnarray}}

\def\beq{\begin{equation}}
\def\eeq{\end{equation}}
\def\st{Stueckelberg\ }

\def\s1{$s_{\alpha}$}
\def\s2{$s_{\gamma}$}
\def\s3{$s_{\delta}$}
\def\c1{$c_{\alpha}$}
\def\c2{$c_{\gamma}$}
\def\c3{$c_{\delta}$}
\def\s{Stueckelberg~}

\newcommand{\mathsym}[1]{{}}

\def\beq{\begin{equation}}
\def\eeq{\end{equation}}
\def\beqn{\begin{eqnarray}}
\def\eeqn{\end{eqnarray}}
\def\st{Stueckelberg\ }

\def\s1{$s_{\alpha}$}
\def\s2{$s_{\gamma}$}
\def\s3{$s_{\delta}$}
\def\c1{$c_{\alpha}$}
\def\c2{$c_{\gamma}$}
\def\c3{$c_{\delta}$}
\def\s{Stueckelberg~}

\def \cha{\widetilde{\chi}^{\pm}_1}

\def \na{\widetilde{\chi}^{0}_1}
\def \nb{\widetilde{\chi}^{0}_2}
\def \nc{\widetilde{\chi}^{0}_3}

\def \g{\widetilde{g}}

\def \ta{\widetilde{t}_1}

\def \sta{\widetilde{\tau}_1}
\def \stb{\widetilde{\tau}_2}
\def \smr{\widetilde{\mu}_R}
\def \ser{\widetilde{e}_R}

\def \slr{\widetilde{l}_R}

\def \snl{\widetilde{\nu}_{\tau}}
\def \snm{\widetilde{\nu}_{\mu}}

\def \hc{H^{\pm}}

\def\beq{\begin{equation}}
\def\be{\begin{equation}}
\def\beqn{\begin{eqnarray}}
\def\ee{\end{equation}}
\def\eeq{\end{equation}}
\def\eeqn{\end{eqnarray}}

\def\beq{\begin{equation}}
\def\be{\begin{equation}}
\def\beqn{\begin{eqnarray}}
\def\ee{\end{equation}}
\def\eeq{\end{equation}}
\def\eeqn{\end{eqnarray}}

\def \cha{\widetilde{\chi}^{\pm}_1}

\def \na{\widetilde{\chi}^{0}_1}
\def \nb{\widetilde{\chi}^{0}_2}
\def \nc{\widetilde{\chi}^{0}_3}

\def \g{\widetilde{g}}

\def \ta{\widetilde{t}_1}

\def \sta{\widetilde{\tau}_1}
\def \stb{\widetilde{\tau}_2}

\def \smr{\widetilde{\mu}_R}
\def \ser{\widetilde{e}_R}

\def \slr{\widetilde{l}_R}

\def \snl{\widetilde{\nu}_{\tau}}
\def \snm{\widetilde{\nu}_{\mu}}

\def \hc{H^{\pm}}

\def\co{coannihilation~}

\def \cha{\widetilde{\chi}^{\pm}_1}

\def \nb{\widetilde{\chi}^{0}_2}
\def \nc{\widetilde{\chi}^{0}_3}

\def \g{\tilde{g}}

\def \ta{\widetilde{t}_1}

\def \sta{\widetilde{\tau}_1}
\def \stb{\widetilde{\tau}_2}
\def \smr{\widetilde{\mu}_R}
\def \ser{\widetilde{e}_R}

\def \slr{\widetilde{l}_R}

\def \snl{\widetilde{\nu}_{\tau}}
\def \snm{\widetilde{\nu}_{\mu}}

\def \hc{H^{\pm}}

\def\.4{\vspace{-.5cm}}

\def\c{GNLSP$_{\rm C}~$}

\usepackage{amssymb,graphics,graphicx, epsfig}
\usepackage{amsfonts}
\usepackage{hyperref}
\usepackage{multicol}
\usepackage{bm}
\usepackage{hyperref}
\title{\begin{flushright}
{\small \bf MCTP-09-43}
\end{flushright}
\bf Superparticle Signatures: from PAMELA to the LHC } 
\author{
\bf Daniel Feldman \\
Michigan Center for Theoretical Physics,\\
 University of Michigan, Ann Arbor
}

\begin{document}

\begin{abstract}
Signatures of soft supersymmetry breaking at the CERN LHC and in dark matter experiments are discussed
with focus drawn to light superparticles, and in particular light gauginos and their discovery prospects. Connected
to the above is the recent PAMELA positron anomaly and its implications for signatures of SUSY in early runs
at the Large Hadron Collider. Other new possibilities for physics beyond the Standard Model are also briefly
discussed.
\vspace{-.2cm}
\end{abstract}

\maketitle

\vspace{-.4cm}
\section{Dual Probes of SUSY}
\vspace{-.2cm}
We review here  testable predictions of 
high scale models with universal and non universal
soft [supersymmetry]\footnote{For recent clear reviews see \cite{Nath}} breaking within the framework of
applied supergravity (SUGRA) and effective models of string theories with D-branes supporting chiral gauge theories 
(for recent related reviews see \cite{NP,Blumenhagen:2006ci}).  Common to all these models are the 
ingredients needed for working in the predictive SUGRA framework, namely:
(a) an effective  K\"{a}hler metric which generally depends on moduli, 
(b) a gauge kinetic function also dependent on such scalars, 
and (c) a superpotential  comprised of visible and hidden
sector fields and a bilinear term for the Higgses. 

An important set of predictions of the models discussed here
is that they can offer the possibility of a relatively light
gluino and electroweak gauginos with dark matter which is naturally Majorana.
Thus  the confluence
of LHC signatures and signatures of dark matter  play a central role in understanding the predictions of 
the above models. This connection is illuminated through knowledge of the possible sparticle mass hierarchies
that can arise \cite{Feldman:2007zn}.
These mass hierarchies include the possibility of light scalars. However, naturalness/radiative electroweak symmetry breaking (REWSB)
tend to point us to light gauginos and heavy squarks which generally occur on the upper Hyperbolic Branch (HB) of REWSB \cite{Chan:1997bi} often
referred to as the focus point (FP) region. This region naturally arises in the minimal supergravity framework \cite{Chamseddine:1982jx}  and its extensions which are typically perturbations around universality.

Towards the end of this overview we 
will further discuss the connection between dark matter and the LHC,
but more specifically in terms of the link between the WMAP data and 
the recent PAMELA positron excess \cite{Spergel:2006hy}.
Should the PAMELA anomaly be attributed
to SUSY dark matter, the eigen-composition of the LSP plays a very relevant role. 
In addition, the composition of the LSP 
has important implications for collider signatures, and thus
a direct bearing on the discovery prospects of SUSY at LHC,
as well as the very nature of how dark matter was produced in the early universe.
 
\vspace{-.3cm}
\section{Resolving the Sparticle Landscape}
\vspace{-.2cm}
We begin with the Sparticle Landscape\cite{Feldman:2007zn}.
Of the 32 massive particles predicted in the MSSM, the
 number of ways in which the sparticle masses can stack up in their
mass hierarchy is a priori undetermined unless an underlying framework
is specified.  Thus, if the 
 32 masses are treated as essentially all independent, then aside
from sum rules  on
the Higgs, sfermions, and gaugino masses, and  without
imposition of any  phenomenological constraints, the number of
hierarchical patterns for all 32 sparticles could be as many as
$O(10^{25-28})$ or larger. 
One may compare this with the landscape of string vacua in
type II strings which lead to $O(10^{1000})$ possibilities.
However, the number of superparticle mass hierarchies 
is reduced enormously in predictive frameworks
such as supergravity models with REWSB.

Upon carrying out a
mapping of the parameter space of the minimal SUGRA framework  for
the first four particles (discounting the lightest Higgs whose
mass is constrained over a $\sim$ 25 GeV range)  we find 22 possible 
mass patterns consistent with all known collider and cosmological constraints.
We label these as  mSUGRA
pattern 1 (mSP1) through mSUGRA pattern 22 (mSP22), the first 16 arising
for $\mu >0$ and the remaining for $\mu <0$ \cite{Feldman:2007zn}.
In Table(\ref{msptable})
we exhibit these mass orderings.
The groupings may be considered more simply in
terms of the NLSP; thus there are Chargino Patterns (CPs) (mSP 1-4), Stau Patterns 
(SUPs) (mSP 5-10,17-19),
Stop Patterns (SOPs) (mSP 11-13,20,21), Higgs Patterns (HPs) (mSP 14-16), and an isolated Neutralino Pattern (mSP 22).

\begin{table}[h]
    \begin{center}
\begin{tabular}{|l||l|c|}
\hline\hline
mSP&     Mass Pattern & $\mu$
\\\hline\hline
mSP1    &   $\na$   $<$ $\cha$  $<$ $\nb$   $<$ $\nc$   & $\mu_{\pm}$    \cr
mSP2    &   $\na$   $<$ $\cha$  $<$ $\nb$   $<$ $A/H$  & $\mu_{\pm}$    \cr
mSP3    &   $\na$   $<$ $\cha$  $<$ $\nb$ $<$ $\sta$    & $\mu_{\pm}$    \cr
mSP4    &   $\na$   $<$ $\cha$ $<$ $\nb$   $<$ $\g$    & $\mu_{\pm}$    \cr
\hline
mSP5    &   $\na$ $<$ $\sta$  $<$ $\slr$  $<$ $\snl$      & $\mu_{\pm}$    \cr
mSP6 &   $\na$   $<$ $\sta$  $<$ $\cha$  $<$ $\nb$     & $\mu_{\pm}$    \cr
mSP7    &   $\na$   $<$ $\sta$  $<$ $\slr$  $<$ $\cha$  & $\mu_{\pm}$    \cr
mSP8    &   $\na$ $<$ $\sta$  $<$ $A\sim H$            & $\mu_{\pm}$    \cr
mSP9    &   $\na$   $<$ $\sta$  $<$ $\slr$ $<$ $A/H$    & $\mu_{\pm}$    \cr
mSP10   &   $\na$   $<$ $\sta$ $<$ $\ta$ $<$ $\slr$     & $\mu_{+}$    \cr
 \hline
mSP11   &   $\na$ $<$ $\ta$ $<$ $\cha$  $<$ $\nb$      & $\mu_{\pm}$    \cr
mSP12 &   $\na$ $<$ $\ta$   $<$ $\sta$ $<$ $\cha$   & $\mu_{\pm}$    \cr
mSP13   & $\na$   $<$ $\ta$ $<$ $\sta$  $<$ $\slr$      & $\mu_{\pm}$    \cr
\hline
mSP14   &   $\na$   $<$  $A\sim H$ $<$ $\hc$       & $\mu_{+}$    \cr
mSP15   &   $\na$   $<$ $ A\sim H$ $<$ $\cha$    & $\mu_{+}$    \cr
mSP16   &   $\na$   $<$ $A\sim H$ $<$ $\sta$         & $\mu_{+}$    \cr
\hline
mSP17   &   $\na$   $<$ $\sta$ $<$ $\nb$ $<$ $\cha$     & $\mu_{-}$    \cr
mSP18   &  $\na$   $<$ $\sta$  $<$ $\slr$  $<$ $\ta$    & $\mu_{-}$    \cr
mSP19   &  $\na$ $<$ $\sta$ $<$ $\ta$   $<$ $\cha$    & $\mu_{-}$    \cr
 \hline
mSP20  & $\na$ $<$ $\ta$   $<$ $\nb$   $<$ $\cha$    & $\mu_{-}$    \cr
mSP21   & $\na$   $<$ $\ta$   $<$ $\sta$  $<$ $\nb$      & $\mu_{-}$    \cr
\hline
mSP22   & $\na$   $<$ $\nb$   $<$ $\cha$  $<$ $\g$   & $\mu_{-}$    \cr
\hline\hline
 \end{tabular}
\caption{\small The Sparticle Landscape of Mass Hierarchies in mSUGRA.
 In patterns mSP14,15,16  the LSP
$\tilde \chi_1^0$  and the Higgs bosons $(A,H)$ can actually 
switch  their order. (From  Refs.(1,3) of \cite{Feldman:2007zn}.)
}
\label{msptable}
\end{center}
 \end{table}

By extending the minimal framework to include 
larger landscapes in SUGRA models with
non universalities (NUSUGRA) and in  D-brane models we find new mass hierarchies.
For the NUSUGRA cases, motivated by flavour changing neutral current constraints, considered are the following three
possibilities  for the soft parameters at the GUT scale:   (i)   the Higgs
sector (NUH):  $M_{H_u,H_d} = m_0(1+\delta_{H_{(u,d)}})$ (ii) the third  generation squark sector (NU3):
$M_{q3} = m_0(1+\delta_{q3}),~M_{u3,d3} =  m_0(1+\delta_{tbR})$, and
(iii)  the gaugino sector (NUG): $M_{1,2,3} = m_{1/2}\{1,(1+\delta_{2}),(1+\delta_{3})\}$.

\begin{table}[h]
    \begin{center}
\begin{tabular}{|l|l|c|c|}
\hline\hline NUSP &   Mass Pattern  &     Model\\\hline\hline
NUSP1   &   $\na$   $<$ $\cha$  $<$ $\nb$   $<$ $\ta$       & NU3,NUG  \cr
NUSP2   &   $\na$   $<$ $\cha$  $<$ $A\sim H$               & NU3        \cr
NUSP3   &   $\na$   $<$ $\cha$  $<$ $\sta$  $<$ $\nb$       & NUG \cr
NUSP4   &   $\na$   $<$ $\cha$  $<$ $\sta$  $<$ $\slr$     & NUG \cr
\hline
NUSP5   &   $\na$   $<$ $\sta$  $<$ $\snl$  $<$ $\stb$       & NU3     \cr
NUSP6  &   $\na$   $<$ $\sta$ $<$ $\snl$  $<$ $\cha$        & NU3  \cr
NUSP7   &   $\na$ $<$ $\sta$  $<$ $\ta$   $<$ $A/H$         & NUG \cr
NUSP8   &   $\na$   $<$ $\sta$  $<$ $\slr$  $<$ $\snm$      & NUG   \cr
NUSP9   &   $\na$   $<$ $\sta$  $<$ $\cha$  $<$ $\slr$       & NUG  \cr
\hline
NUSP10  &   $\na$   $<$ $\ta$   $<$ $\g$    $<$ $\cha$         & NUG  \cr
NUSP11  &   $\na$   $<$ $\ta$   $<$ $A\sim H$                      & NUG  \cr
\hline
NUSP12  &   $\na$   $<$ $A\sim H$   $<$ $\g$                    & NUG  \cr
\hline
NUSP13  &   $\na$   $<$ $\g$    $<$ $\cha$ $<$ $\nb$         & NUG  \cr
NUSP14  &   $\na$   $<$ $\g$    $<$ $\ta$ $<$ $\cha$            & NUG  \cr
NUSP15  &   $\na$   $<$ $\g$    $<$ $A\sim H$               & NUG  \cr
\hline
{\rm DBSP1} & $\na < \sta <  \snl    <A/H$   & DB  \cr
{\rm DBSP2} & $\na < \sta  <   \snl  <\slr$  & DB \cr
{\rm DBSP3} & $\na < \sta < \snl  <\snm $   & DB \cr
{\rm DBSP4} &  $\na  < \ta  < \sta  <\snl $  & DB  \cr
{\rm DBSP5} & $\na < \snl  < \sta <\snm $   & DB  \cr
{\rm DBSP6} & $\na <  \snl  < \sta <\cha $  & DB  \cr
\hline\hline
\end{tabular}
\end{center}
\label{patternstable2}
\caption{ New sparticle mass hiearchies above and beyond those
in the minimal framework in NUSUGRA and the D-brane model. Here one has  gluino NLSP (GNLSP) patterns
as well as the possibility of a light sneutrino of the third generation as the NLSP. (From  Refs.(2,3) of \cite{Feldman:2007zn}.)}
 \end{table}


The landscape analysis reveals a collection of
new  mass hierarchies 
not discussed in the literature; even in the minimal model.
Indeed most of the mSP patterns do not appear in previous works. All
of the Snowmass mSUGRA points (labeled SPS)
 are only of types mSP(1,3,5,7).
Regarding the Post-WMAP points, these are predominantly
stau-coannihilation models, and one model which resides on the Hyperbolic Branch.
The CMS benchmarks classified as  (Low/High) Mass (LM)/(HM)
  \cite{Ball:2007zza} do a better job of representing mSP1 which is
one of several dominant patterns, however, again the mapping only covers
mSP(1,3,5,7) (see \cite{Feldman:2007zn} for details).
 One such example of a missing class of mass hierarchies are the 
 Higgs Patterns mSP(14-16) (HPs)\cite{Feldman:2007zn} which occur for large $\tan \beta$ in mSUGRA (but occur
more generally at lower values in NUSUGRA)
where the CP odd/heavier CP even Higgses are in fact the lightest particles
beyond the LSP neutralino and in some cases they can be even lighter (we remind
the reader that the Higgses are R-Parity even). The HPs typically occur in the bulk
and will be discussed further in the next section.
For the large $m_0$
region with lower values of $m_{1/2}$ the mass hierarchy is dominantly composed of mSP1 
located in the HB/FP region where the LSP has a strong Higgsino component,
 while there is also a bino branch of mSP1 in which the the lightest
chargino and next heavier neutralino are very close in mass, really effectively
degenerate. One also observes several more CPs where
the NLSP is a chargino (mSP1-mSP4), and where mSP4 is rather special with a very
light neutralino mass less than $\sim$ 55 GeV. Bino dominated patterns include, for example, the
(SUPs) mSP(5-10) and (SOPs) mSP(11-13).  A
 large density of the model points also occur in the vicinity of a Higgs pole \cite{Nath:1992ty}. Many models generally arise also from
coannihilations \cite{Griest:1990kh}, for example, mSP(5,7)  and mSP(11) arise respectively from $\sta-\na$ and $\ta-\na$ \co \cite{Ellis:1999mm}, however
many of the new patterns actually arise from a  large mixture of thermal annihilations.

For the case
of non universal SUGRA models,  15 new mass patterns are uncovered
labeled NUSP(1-15). In a class of models based on D-branes (to be discussed shortly),
 6 more new patterns labeled as DBSP(1-6) are also revealed. 
A complete set of specific benchmarks for each pattern can be found in \cite{Feldman:2007zn}. 
There are also patterns where the gluino is light and is the NLSP ($\equiv \rm GNLSP$ discussed in what follows)
and where the sneutrino of the third generation is light and is the NLSP occurring
in the NU and DB cases respectively as illustrated in Table(\ref{patternstable2}).

\vspace{-.4cm}
\section{Light Higgs Bosons and D-branes} 
\vspace{-.2cm}
\label{db}

\begin{figure*}[t]
  \begin{center}
\includegraphics[width=5cm,height=5cm]{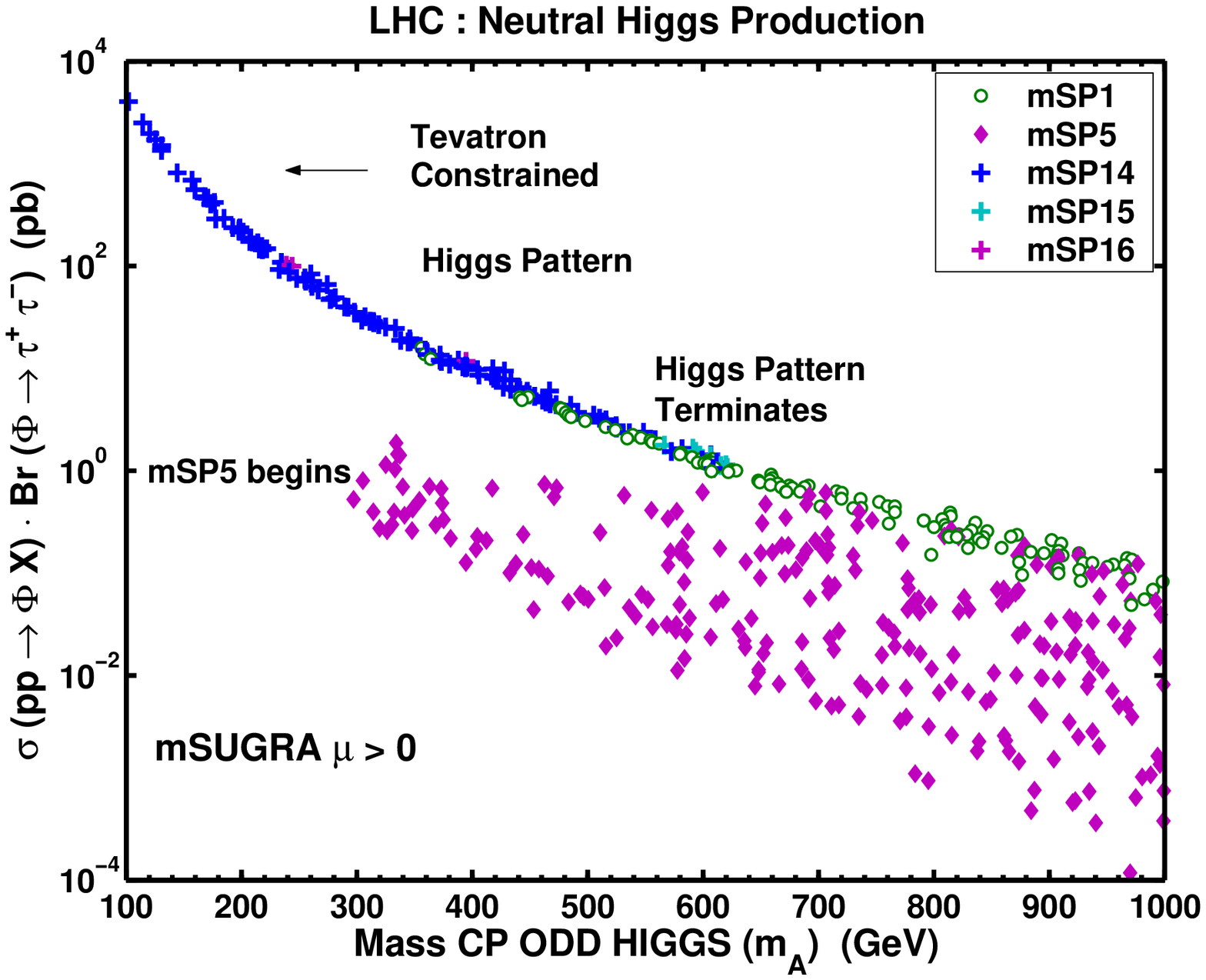}
\includegraphics[width=5cm,height=5cm]{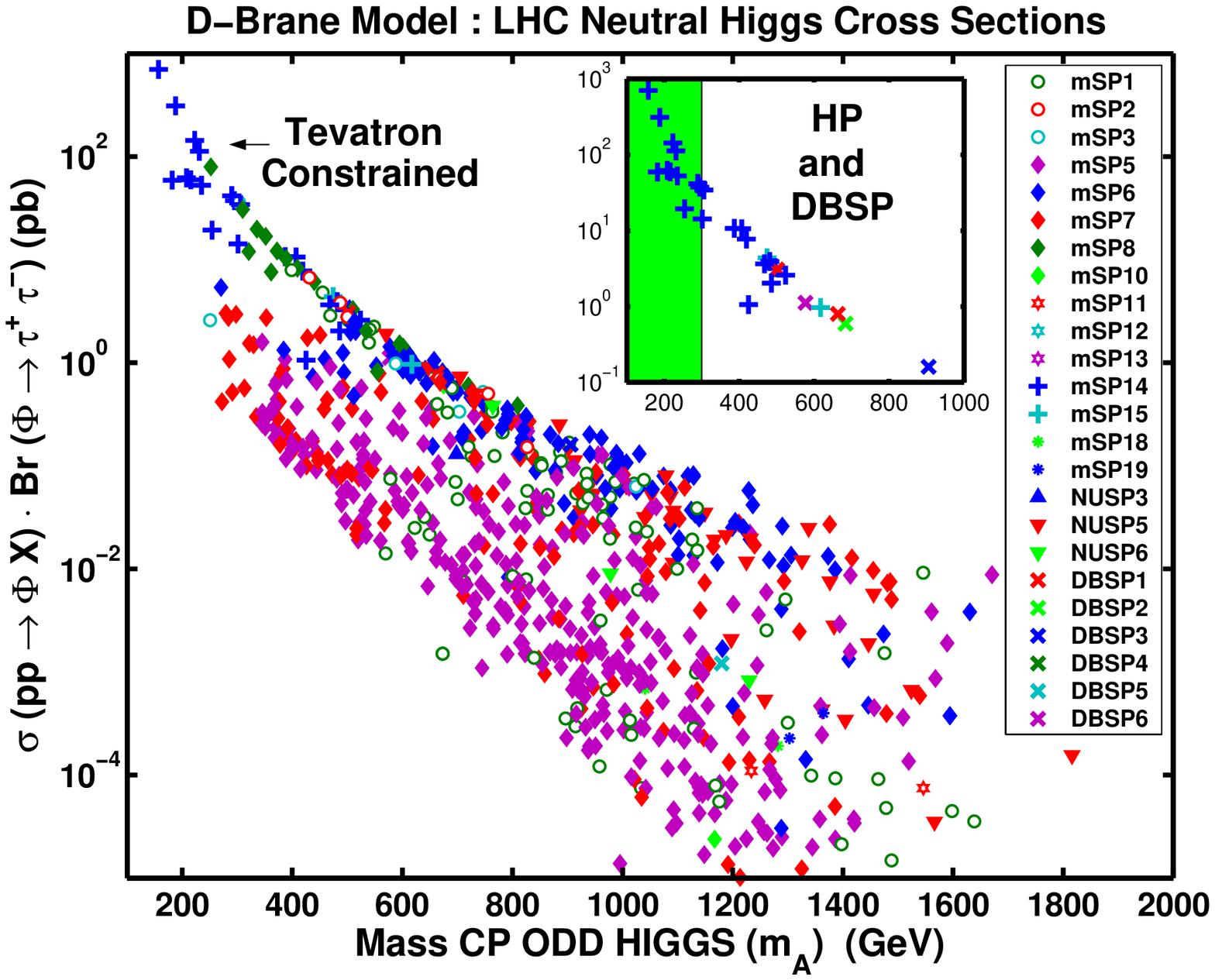}
\includegraphics[width=5cm,height=5cm]{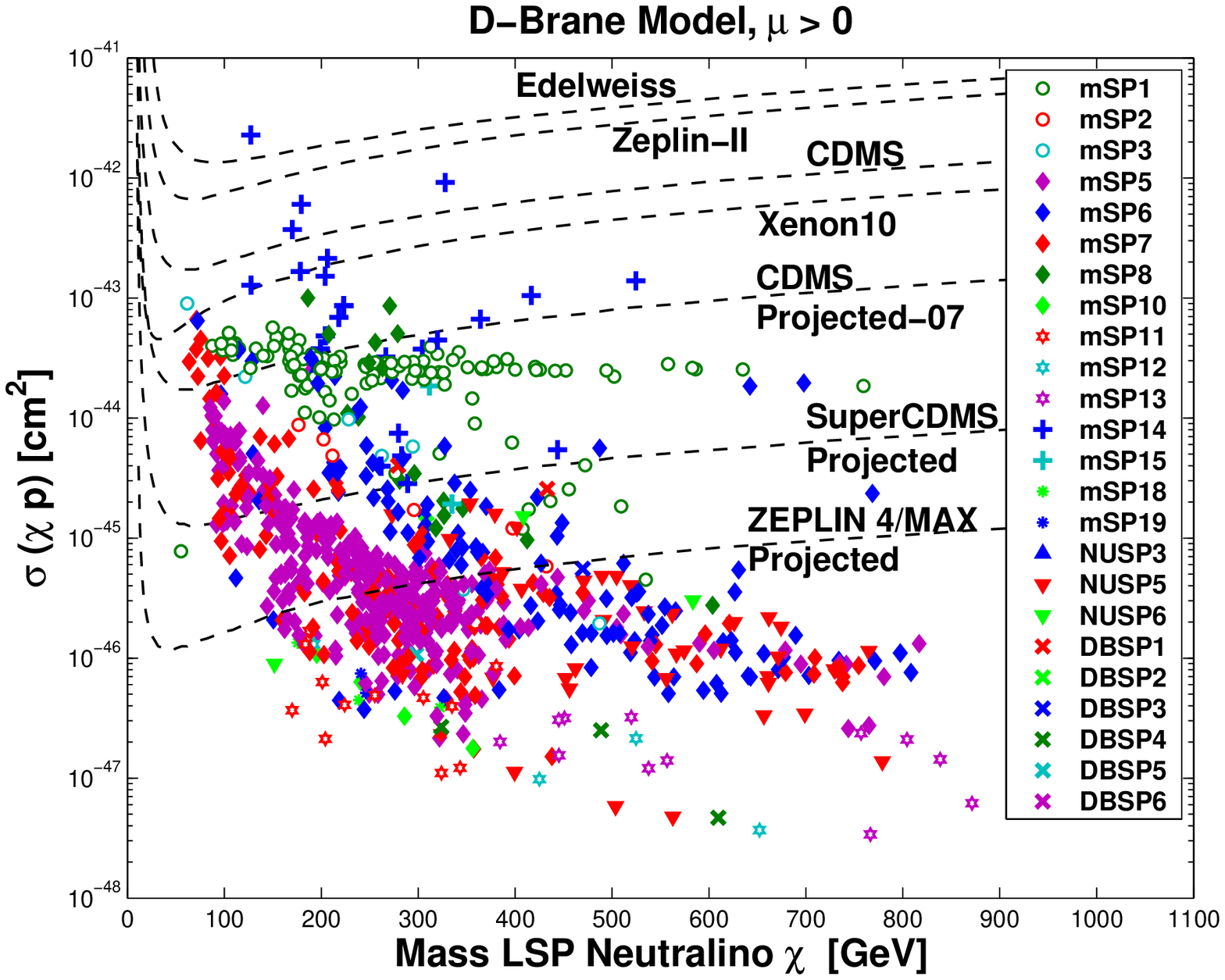}
\caption{
The neutral Higgs  production cross section at the LHC in the $2 \tau $ mode as a
function of the CP odd  Higgs mass $m_A$ for several patterns in mSUGRA (left panel) and 
(middle) all patterns arising in the D-brane model. The dark matter direct detection
signature space for the D-brane model is also exhibited (right panel). (From  Ref.(2) of \cite{Feldman:2007zn}.)}
\label{KN}
  \end{center}
\end{figure*}
Progress in D-brane model building   (see e.g., 
 \cite{Antoniadis:2000ena,Ibanez:2001nd,Cvetic:2001nr,NT,Cremades:2003qj,Kors:2003wf,dsoft2,Font:2004cx,Kors:2004hz,Bertolini:2005qh,Blumenhagen:2006ci})
has lead to testable predictions of related models which support chiral matter
 \cite{Kane:2004hm,Kane:2006yi,Feldman:2007zn,Chen:2007zu}.  
   Here we discuss 
light Higgses and dark matter in the context of SUGRA and D-branes \cite{Feldman:2007zn}. 
We first briefly summarize the
 model class studied in 
 \cite{Kors:2003wf} which offers a concrete early example of an 
 effective string model where the soft terms can be computed. The
model employs 
toroidal  orbifold compactifications based on  ${\cal
T}^6/\mathbb{Z}_2\times \mathbb{Z}_2$ where ${\cal T}^6$ is taken
to be  a product of 3  ${\cal T}^2$ tori.   This model has a moduli
sector consisting of volume moduli $t_m$,  shape moduli $u_m$  $(m
=1,2,3)$  and the axion-dilaton field $s$. Of special interest here
is the K\"{a}hler metric of the $m^{\rm th}$ component of open strings
which are split between common brane stacks $[a,a]$ and twisted open
strings connecting different brane
stacks $[a,b]$. The K\"{a}hler metric is deduced from dimensional reduction
and $T-$duality generalizing the previous known result for the heterotic string \cite{Kors:2003wf}. 
The soft scalars are simple functions
of the graviton mass, the stack angle, and moduli VEVs and are given in full in  \cite{Kors:2003wf}.
Specifically the parameter space consists
of the gravitino mass $m_{3/2}$, the gaugino mass $m_{1/2}$, the
tri-linear coupling $A_0$ (which is in general non-vanishing),  $\tan\beta$, sign($\mu$), the stack angle $\alpha$
($0\leqslant \alpha \leqslant \frac{1}{2}$), and the $F$ term  VEVs through the Goldstino
angle $\theta$, and complex modular parameters 
 $\Theta_{t_i}$, $\Theta_{u_i}$  $(i=1,2,3)$ . The latter parametrize the directionality of the Goldstino
 angles in the modular space and enter in the VEV of the $F$-term which are functions of the real part of the moduli.

From the analysis of the D-brane model it is found  that the mass hierarchies are
dominated by
the mSPs with only six new patterns (at isolated points) emerging. 
Of specific interest is that all the HPs (mSP14-mSP16) are
seen to emerge in good abundance. Given in Fig.(\ref{KN}) (left panel) is the analysis of the neutral Higgs  [$\Phi = h,H,A$] production cross sections
at the LHC in the $2\tau$ mode for several patterns in the minimal SUGRA model with the dominant 
contributions to the cross section entering from gluon fusion and bottom quark annihilation processes. Similarly, in  the  middle panel of Fig.(\ref{KN}) are the Higgs production cross sections in the D-brane model, where all patterns seen in the D-brane model are exhibited. 
 The
analysis shows that the HPs dominate the Higgs production
cross sections.  One also finds that the $B_s\to \mu^+\mu^-$ predictions constrain
the HPs in this model\cite{Feldman:2007zn}.
The  spin  independent proton-LSP cross  sections are  given  in the right panel of Fig.(\ref{KN}).  Here also one finds  that the Higgs Patterns  typically give the largest
scalar cross sections followed by the Chargino Patterns (mSP1-mSP3)
 and then by  the Stau Patterns (which are dense in this model).
Further, one finds a  Wall  of  Chargino
Patterns developing which enhances the discovery potential of 
these CPs (see also the middle panel of Fig.(\ref{bigpic})). This Wall is 
consequence of the larger Higgsino component of mSP1 which is also
found in the DB model class \cite{Feldman:2007zn}.

\vspace{-.35cm}
\section{Compressed Spectra in Intersecting D-brane Models}
\vspace{-.3cm}
\label{db2}
We consider next another  
class of  D-brane models, 
motivated by the recent works of  
\cite{Cremades:2003qj,dsoft2,Font:2004cx,Bertolini:2005qh}.
The specific class of models we consider is with $u$ moduli breaking.
The model consists of
a chiral particle spectrum arising 
from intersecting branes with supporting gauge
groups $SU(3)_C \times SU(2)_L$ and $U(1)_a$,$ U(1)_c$, $U(1)_d$
and $U(1)_Y$. Here, there is an anomalous $U(1)= U(1)_a + U(1)_d$ and the anomaly is cancelled 
 via a Green-Schwarz mechanism (for an overview see \cite{Blumenhagen:2006ci}) giving 
a \st mass to the $U(1)$ gauge boson (for  recent reviews of
the  \st mechanism see  \cite{Feldman:2007nf},\cite{Langacker:2008yv}).
The K\"{a}hler metric for the twisted moduli
arising from strings stretching between stacks $P$ and $Q$
for the BPS $1/4$  sector is taken in the form
similar to \cite{Font:2004cx}, and more specifically
of the form given in \cite{Kane:2004hm} where the gauge kinetic function is also deduced.
The K\"{a}hler metric for  BPS $1/2$ brane configurations in this framework is given in \cite{dsoft2}.
\begin{table}[t]
   \begin{center}
\begin{tabular}{|c||c|c|}
\hline\hline
Sparticle &  D6  & mSUGRA     \\
  type    &   Mass/GeV   &  Mass/GeV 
\\\hline\hline
$m_h$      & 113.9      & 113.6    \cr
 $\na$    & 209.0       & 208.8      \cr
 $\cha$    & 229.1     & 388.6   \cr
 $\nb$    &  229.5      & 388.8   \cr
 $\sta$    & 404.2       & 433.3   \cr
 $\ser,\smr$  & 464.4   & 637.8\cr
 $\sta$      & 547.6       & 929.2\cr
 $\g$      & 760.4        & 1181.4 \cr
$ m_{{\rm max} ={\tilde{s},\tilde{d}}_L}$ &882.2  &  $ m_{{\rm max} ={\tilde{s},\tilde{d}}_L}$ 1210.4 \cr
\hline\hline
 \end{tabular}
\begin{tabular}{|c||c|c|}
\hline\hline
      D6  &  mSUGRA \\
$(\tilde{B},\tilde{W},\tilde{H}_1,\tilde{H}_2)$ &   $(\tilde{B},\tilde{W},\tilde{H}_1,\tilde{H}_2)$     \cr  (0.985,-.133,.104,-.0399)&  (0.994,-.017,.101,-.041)    \cr
 $\sigma^{\rm SI}_{\na p}= 7.4 \times 10^{-9}$ pb    & $\sigma^{\rm SI}_{\na p}= 1.4 \times 10^{-8}$ pb\cr
 $\Omega h^2= 0.099$ co-annih. &  $\Omega h^2= 0.095$ $b \bar b,\tau \bar\tau$\cr
\hline\hline
 \end{tabular}
\caption{Illustration of the compressed spectra in the intersecting D-brane model (D6) and a comparison to a model in mSUGRA.
}
\label{compress}
\end{center}
 \end{table}
In Table(\ref{compress}) a comparison is given  of 2 model points,
one from the D-brane
model (labeled D6) and the other from mSUGRA, both of which sit in mSP3. 
We see from Table(\ref{compress}) that these models have
effectively the same LSP mass and light CP even Higgs mass. Observe also 
that there is a major violation of gaugino mass scaling in the D6 model relative to that of the mSUGRA model.
Quite interestingly, the overall mass scale in the D6 model
is much compressed relative to that of the mSUGRA model. This feature appears rather generic
to the D6 models over the part of parameter space investigated. Thus while the LSP masses
are effectively identical, the NLSP mass in the D6 model is about 160 GeV lighter than in
the mSUGRA case considered. Further,
the $\g$ is several hundred GeV lighter in the D6 case relative to the mSUGRA case 
and the heaviest sparticle in the D6 cases lies lower than the mSUGRA case 
by approximately 300 GeV.
 In Table(\ref{compress}) we see that the D6 LSP has a relatively larger wino 
component, while the LSP in the mSUGRA model is more of a mixed bino-Higgsino but
with a stronger bino component. Both models give a  relic density consistent with WMAP
but do so in very different ways. The compressed
mass scale has implications for collider physics that requires further study.

\vspace{-.3cm}
\section{Dark Matter and the LHC}
\begin{figure*}[]
\centering
\includegraphics[width=6cm,height=5cm]{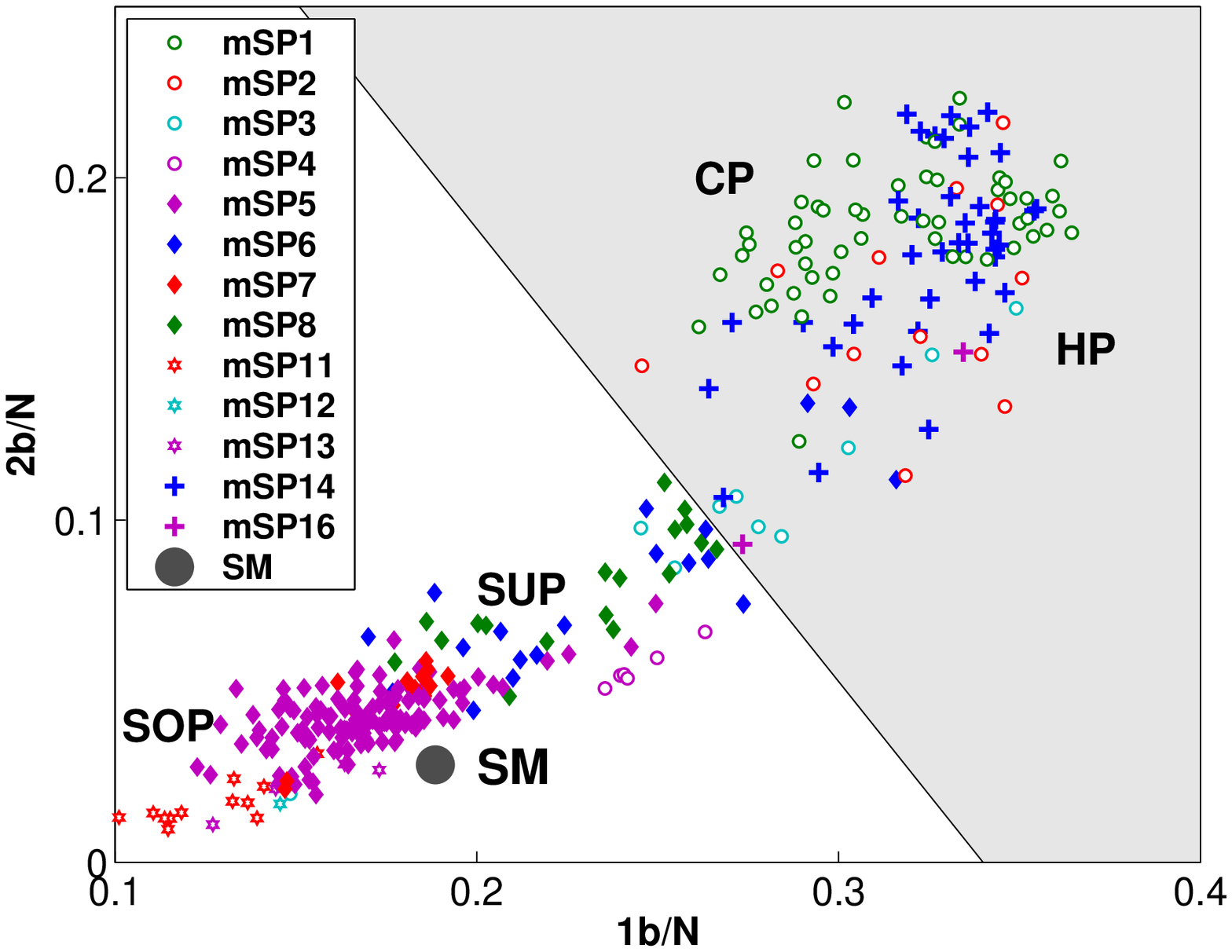}
\includegraphics[width=6cm,height=5cm]{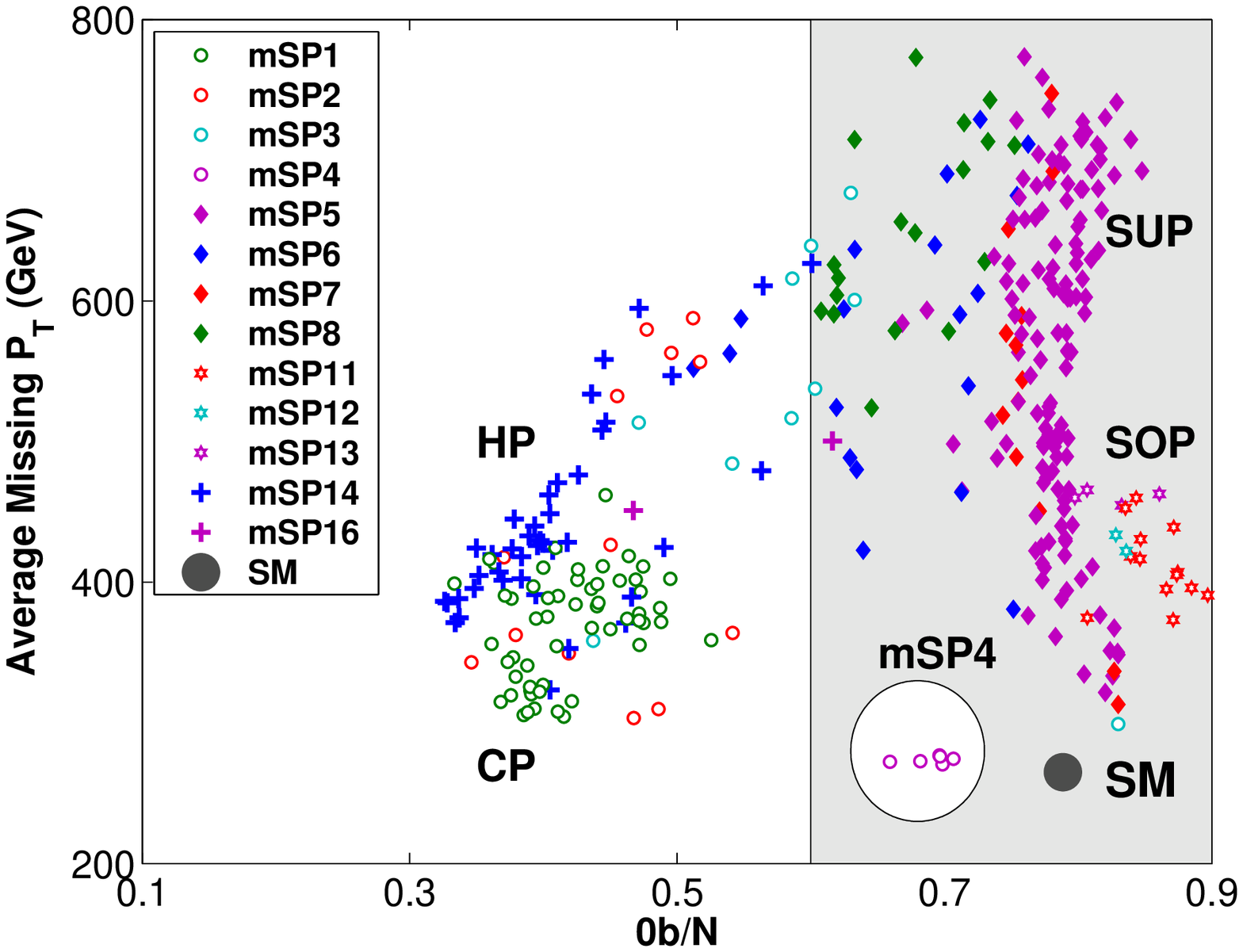}
\includegraphics[width=6cm,height=5cm]{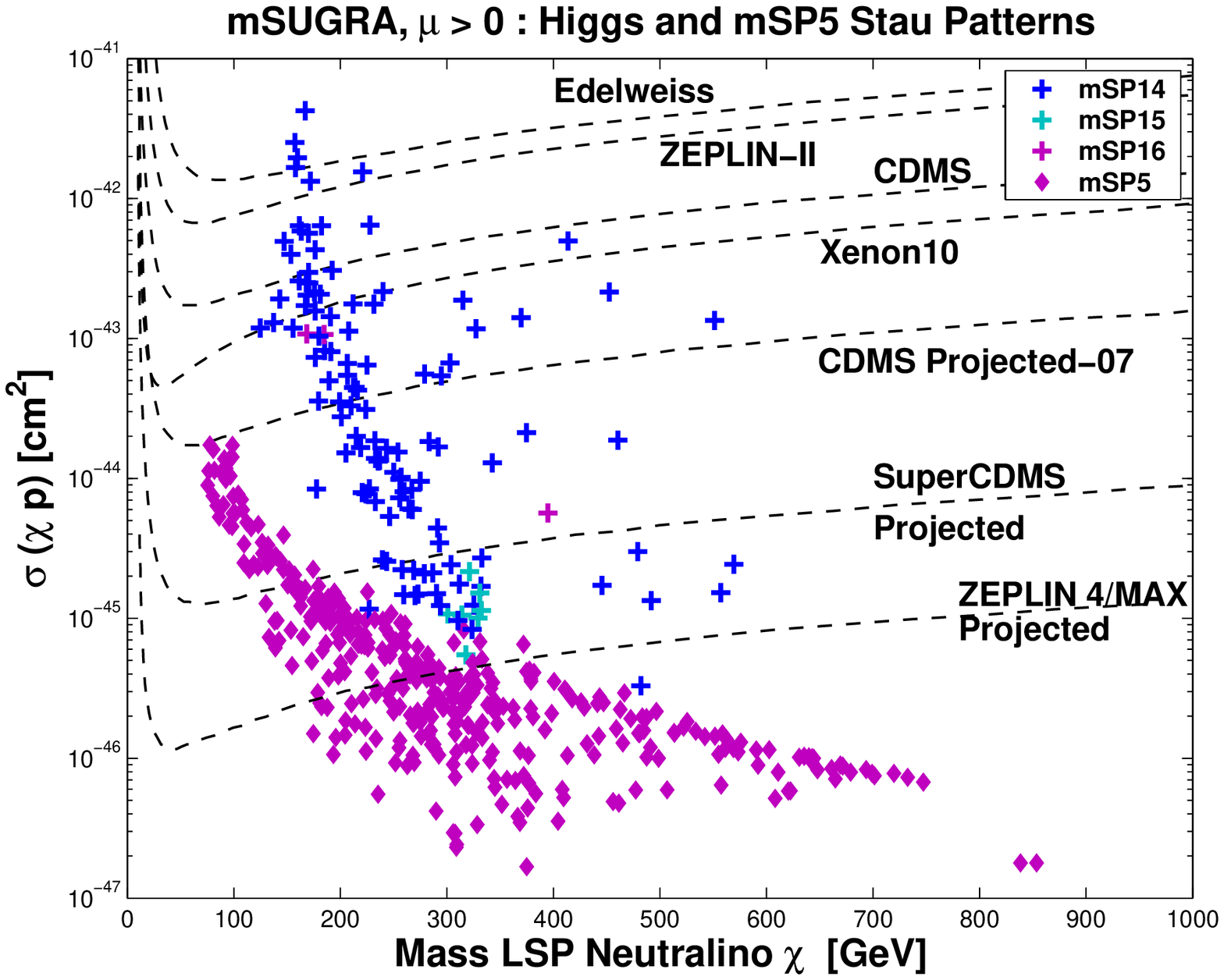}
\includegraphics[width=6cm,height=5cm]{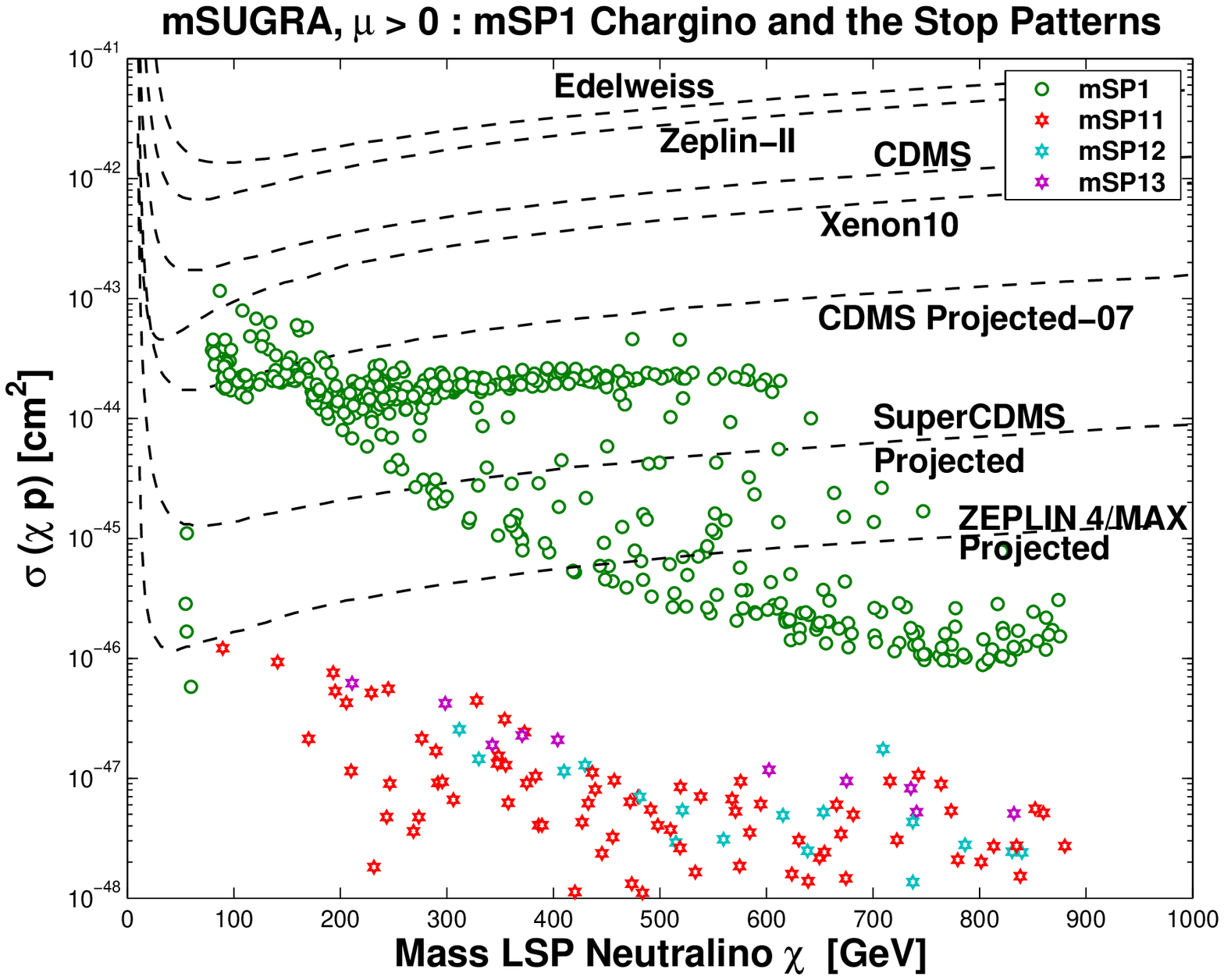}
\includegraphics[width=6cm,height=5cm]{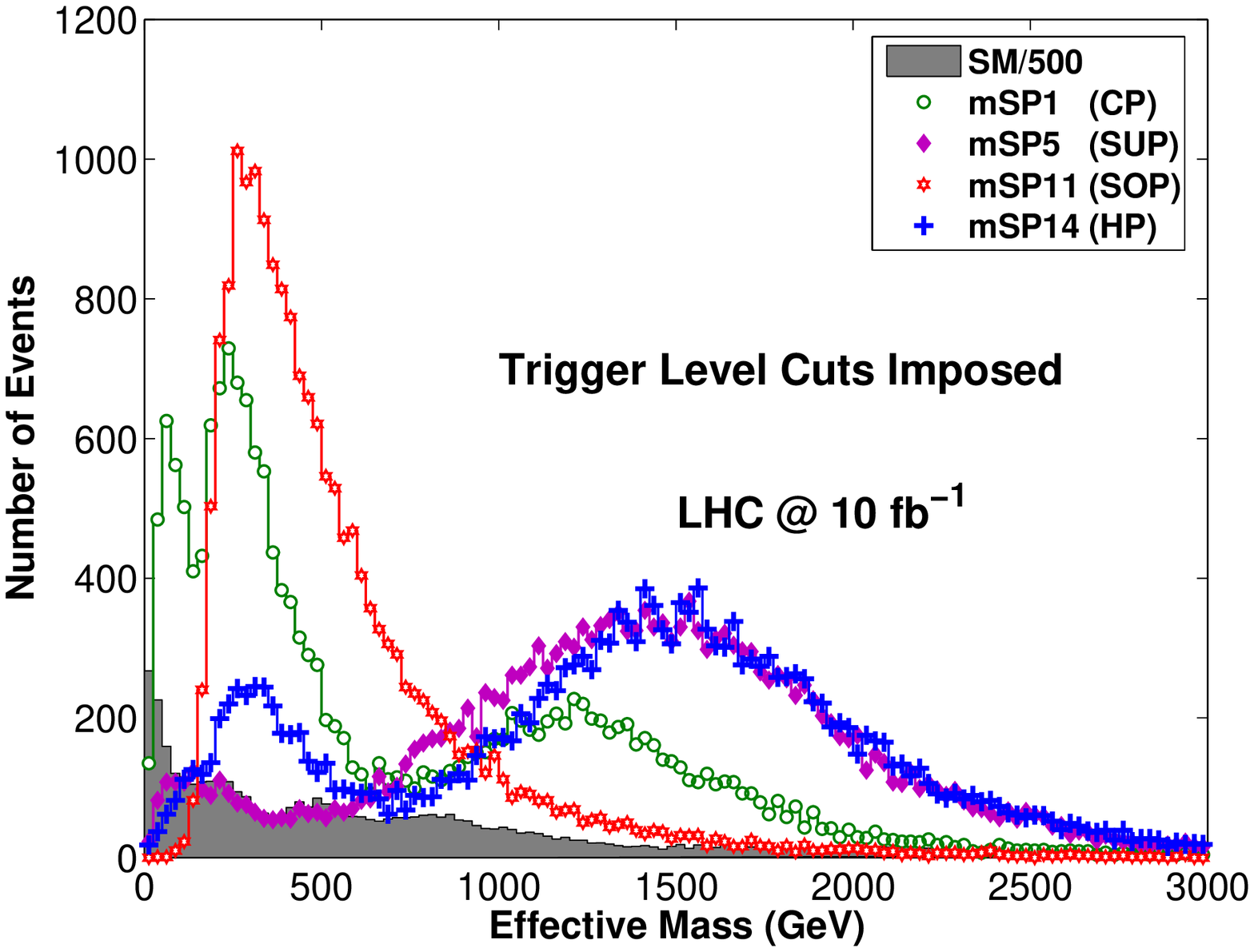}
\includegraphics[width=6cm,height=5cm]{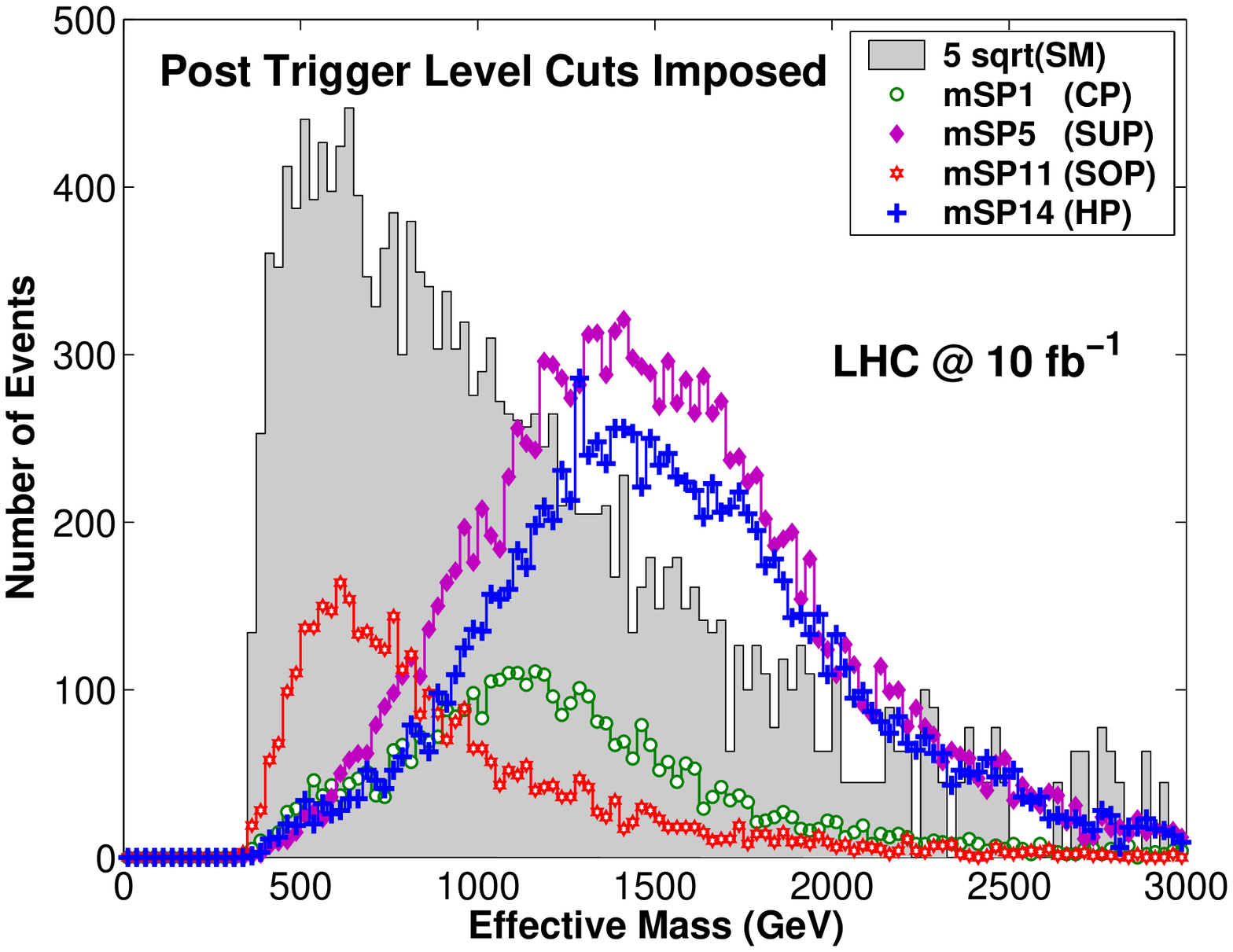}
\caption{\small The top panel
is full simulation of $\sim$ 900 model points for minimal SUGRA at the LHC with 10/fb keeping only
statistically significant model points. The middle panel
has a larger set of data, but corresponds otherwise to same set of models in the top panel
where the low mass points make the cut in the top panel. The bottom panels are the effective mass distribution for 4 benchmarks
with the same cuts at the trigger and  post trigger levels. One observes that the mass hierarchies act as a prism separating
out various hierarchies into imprints on the dual signature spaces. Similar correlations for the non universal cases can be found in 
Refs.(2,3) of \cite{Feldman:2007zn}. 
}
\label{bigpic}
\end{figure*}

\vspace{-.3cm}
We now turn to a very central idea; namely the correlation
of LHC signals with dark matter direct detection signals. 
The correlation is exhibited in Fig.(\ref{bigpic}). The top panel
gives an analysis  at $L=$ 10 fb$^{-1}$ at $\sqrt S = 14 ~\rm TeV$ 
 admitting only model points in the
parameter space that generate at least 500 total SUSY events,
for statistical significance, in the normalized channels
$ [2bjets + jets \geq 2]/N_{\rm SUSY}$ vs $ [1bjet + jets \geq 2]/N_{\rm SUSY}$ 
and average $ P^{\rm miss}_T$ vs  $[0bjets + jets \geq 2]/N_{\rm SUSY}$. 
The middle panels are  4 mSPs shown for illustration (for the full set see the 2nd Ref. of \cite{Feldman:2007zn})
in the
$\sigma^{\rm SI}_{\na p}$ vs LSP signature space. Finally the bottom two 
panels show the effective mass distributions of sample benchmarks for different mSPs.
A large separation among many of the hierarchical patterns
can be seen in  Fig.(\ref{bigpic}). 

Indeed the mass hierarchies act as prism separating the landscape of signatures.
The
top left panel exhibits  separation of CPs and HPs from SOPs
and SUPs, with CPs and HPs occupying one region, and SOPs and SUPs occupy another in
this signature space except for a very small overlap. This can be recast as ``Higgsinos to one side, and binos the to other".
The average missing $P_T$ for each model point vs the fraction of events with $0b$ also shows a separation of the CPs and HPs from SOPs and
SUPs. Further, mSP4 appears isolated residing in its own space.  Even further, the middle panels exhibits the spin independent cross sections
along with direct detection limits and indeed a
remarkable separation of the mass patterns is achieved.

The left bottom panel shows distributions in effective mass where only
trigger level cuts are imposed, while the right panel has post trigger level cuts
imposed and they have been imposed globally for all the models considered (see the 3rd Ref. of \cite{Feldman:2007zn} for these cuts).
For the case when only trigger level cuts are imposed, the SOPs and CPs are highly peaked
at lower values of effective mass, while the  HPs and SUPs are much broader at higher effective mass.
As is apparent, the trigger level cuts can have an enormous
effect on the observability of these signals. 
Thus, imposing the  trigger level cuts
globally on all classes of hierarchical mass patterns may disguise the signal. 
Additionally, the imposed post trigger level cuts  kills the SOP and CP signals, while
the SUPs and HPs signals remain relatively strong. 
Likewise, the missing $P_T$ distributions
for the SOPs and CPs are generally much narrower, while the HPs and SUPs are generally much broader.
We observe these effects
more generally in the minimal SUGRA model. 
The discovery prospects of SUSY will be directly tested by CMS and ATLAS \cite{Ball:2007zza} (see also
\cite{Yamamoto:2007it}) and
prospects continue to be explored at the Tevatron 
(for a sample of recent works by  D0 \& CDF see \cite{MonicaDOnofrio}).  
Triggering for different scenarios is quite important, in particular, as we 
illustrated here, the triggers need to be specialized for different mass hierarchies.
There appears to be some attention now drawn this way for strategies in general \cite{OsamuJinnouchi}.
\vspace{-.4cm}
\section{LHC as a Gluino Factory, GNLSP, and Isomorphic GUT Models }
\vspace{-.2cm}
 \begin{figure*}
 \centering
\includegraphics[width=6.75cm,height=5.75 cm]{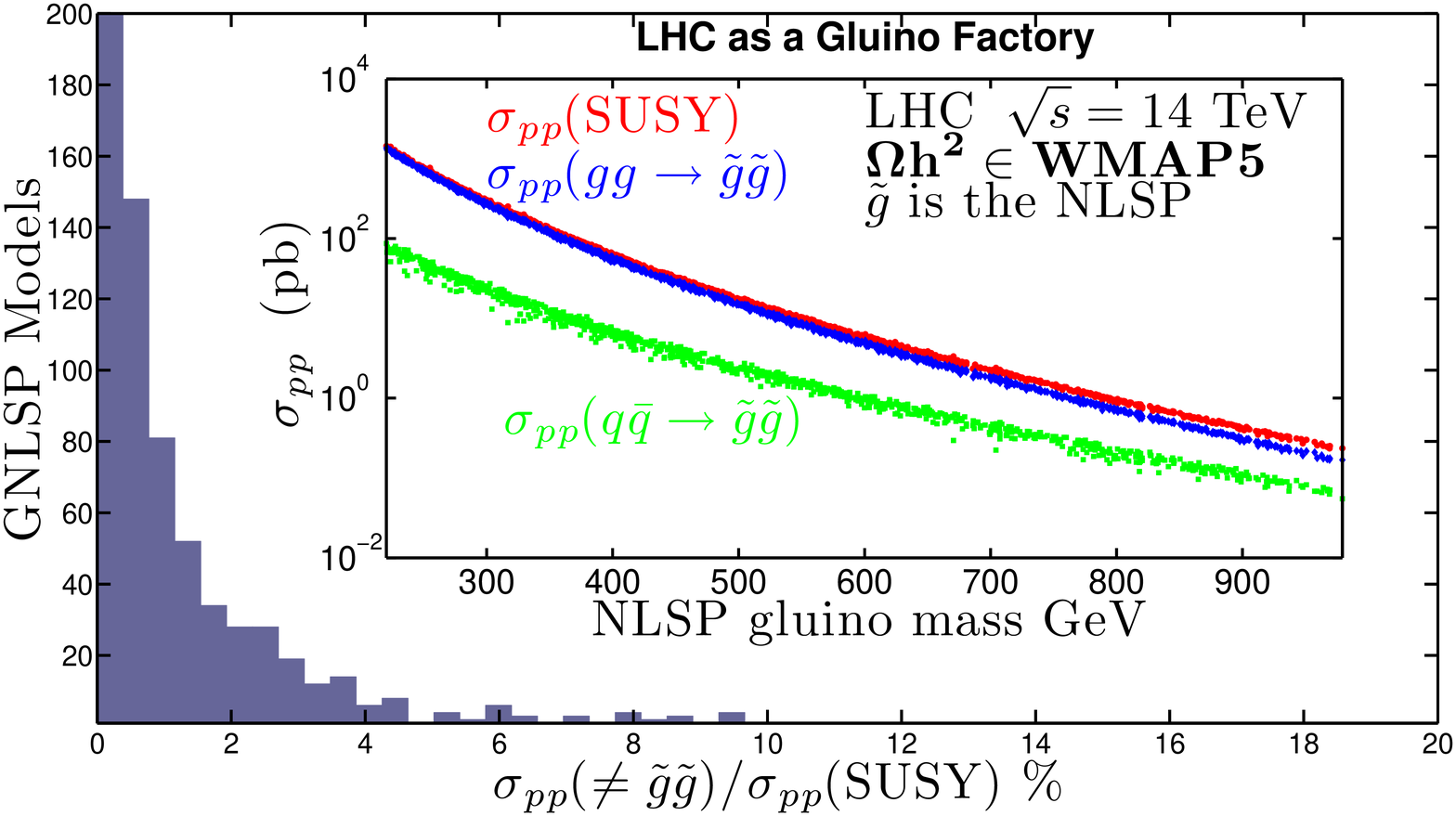}
\includegraphics[width=6.75cm,height=5.75 cm]{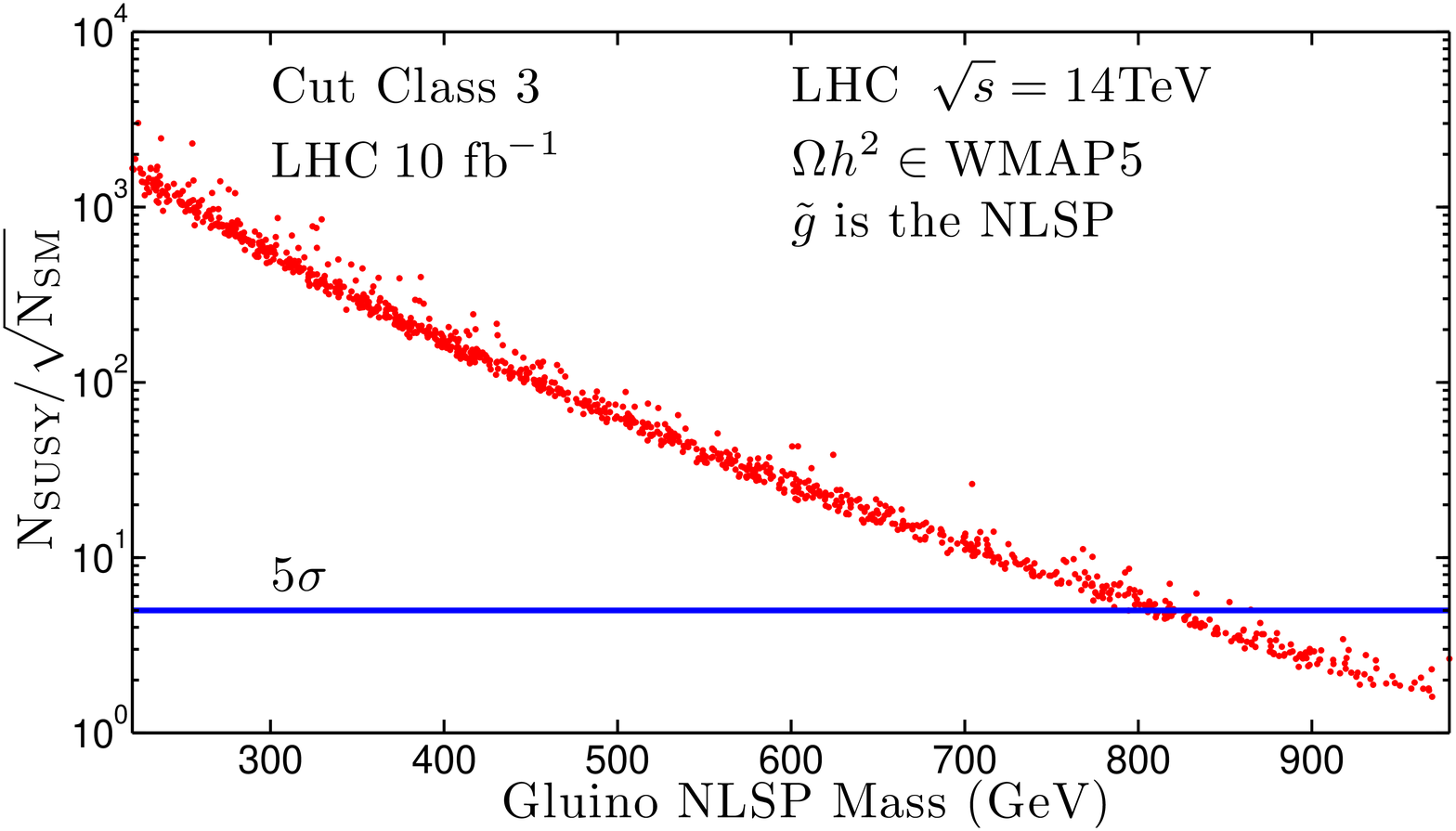}
\vspace{-.3cm}
\caption{Gluino as the NLSP (GNLSP). Through $\g$ coannihilation WMAP relic density constraints are satisfied for $\g$ masses as light as a few hundred GeV and up to a TeV under naturalness assumptions. An overwhelming dominance of $\g$ production is observed \cite{Feldman:2009zc}.
}\label{GF}
\end{figure*}
The possibility of detecting the presence of
a gluino close in mass to the LSP has been raised in \cite{Feldman:2007zn} 
and recently has been studied in \cite{Alwall:2008ve} and in \cite{Feldman:2009zc}. 
This situation arises in supergravity models with  non universalities   in the 
  gaugino mass sector\cite{Feldman:2007zn} and can more specifically come about
   when the GUT symmetry is broken by
a $F$ term in $SU(5), SO(10)$, and $E_6$ models.
In the analysis of \cite{Feldman:2009zc} it is found that a gluino NLSP (GNLSP)  arises 
from such GUT models in the presence of an additional singlet. 
Thus what is minimally  necessary for a GNLSP in these GUT models 
is a combination of GUT symmetry breaking
in the gaugino mass term sector 
with two irreducible representations; a linear combination of 
  a singlet and a non-singlet $F$ term. In this case an  interesting phenomenon arises in that  models with the same 
$r\equiv (M_2-M_1)/(M_3-M_1)$
  can be made isomorphic under redefinitions and scalings in the gaugino sector. 
Therefore, in essence,  models
 with the same value of 
 $r$ would in fact be equivalent, or phenomenologically indistinguishable, when taken  in a linear combination of breakings including singlets. 

One  finds that there are 
several possibilities for which the GNLSP class of models can arise that fall under the isomorphic class of gaugino mass models. Considered more generally are 3 classes of models.
(A)  ${\rm GNLSP_{\rm A}}$ : Here $r$ takes the common value over several models:
$(M_1:M_2:M_3)=$ $(-1/2:-3/2:1)$, $(19/10:5/2:1)$, $(-1/5:-1:1)$;  all of which map into $r=-2/3$.
(B)  ${\rm GNLSP_{\rm B}}$: This is a model specific to $E_6$  with $F$ type breaking with 
 {\bf 2430} plet such that\cite{Martin:2009ad}  $E_6\to SU(6)''\times SU(2)_L( 2430\to (189,1))$
 which gives  $M_1:M_2:M_3=0:0:1$. This model can generate a gluino 
 as the NLSP upon the addition of breaking with a singlet. 
(C)  ${\rm GNLSP_{\rm C}}$: Here $r$ is free and thus defining $r=\delta_2/\delta_3 $, the gaugino masses
at the GUT scale may be parametrized as discussed earlier
and can be varied independently. Model ${\rm GNLSP_{\rm C}}$ contains models ${\rm GNLSP_{\rm A,B}}$ as sub cases.   
For all the three models a GNLSP  requires $\delta_3$ to lie in the range $(-0.9,-0.8)$. 
We also note that from  the analysis of \cite{Martin:2009ad} one can discern several other set of models which have 
 a common value of $r$, i.e other  models such as (A).  Further examples of these isomorphisms are given in  \cite{Feldman:2009zc}
along with generalized sum rules on the gaugino masses.

An interesting property of the GNLSP class of models 
is that the relic density (RD) is controlled by gluino coannihilations \cite{Profumo,Profumo:2005xd},
and one has that 
$\langle \sigma_{\rm eff} v \rangle$  to be integrated  has a cross section well approximated by  \cite{Feldman:2009zc} 
$
\sigma_{\rm eff}\simeq \sigma_{\tilde g \tilde g}\gamma^2_{\na} \left(\gamma^2 + 2\gamma \frac{\sigma_{\na\tilde g}}{\sigma_{\tilde g \tilde g}} + 
\frac{\sigma_{\na\na}}{\sigma_{\tilde g \tilde g} }\right)~,
\label{relic2}
$
where $\gamma =\gamma_{\tilde g}/\gamma_{\na}$ and where $\gamma_i$ are defined in the standard notation of  Ref. \cite{Griest:1990kh}.
Thus for example, a typical set of annihilations {\em for a bino LSP} that contribute to the RD enter with weights
  $ \na \na \to t \bar t (\lesssim 3 \%)$,
 $ \na \na \to \tau^{+}  \tau^{-}  (\sim 1\%)$,
  $ \na \g \to t \bar t (\lesssim 3\%) $,
 $  \g \g \to g g (\sim 50\%)$,
 $  \g \g \to q  \bar q  (\sim 40\%)$.
 We also find  
cases where the GNLSP emerges without significant 
\co which occurs when the LSP 
has a significant Higgsino component \cite{Feldman:2009zc}. In either 
case there is a typical mass splitting of 
 $  \Delta_{\g \na}\equiv(m_{\tilde g}- m_{\na})/m_{\na} \in ( 0.08 - 0.20) $.
 Non-perturbative effects, namely  a Sommerfeld Enhancement of $\langle \sigma_{\rm eff} v \rangle$ , i.e. reduction of $(\Omega h^2)_{\na}$,
requires $\Delta_{\g \na}$ increase by  (2 to 3)\% for $m_{\g} \lesssim \rm TeV$ to maintain  $(\Omega h^2)_{\na} \in$ WMAP \cite{Profumo:2005xd,Feldman:2009zc}. 
We also find, that beyond the compression of the gluino mass  in  GNLSP models, there is a  2nd generation slepton-squark degeneracy (SSD)
and in some cases an inversion where these sleptons  are heavier than the squarks \cite{Feldman:2009zc}.

The dominant SM backgrounds for GNLSP models at the LHC are from QCD, $Z$/$W$+ jets, $b \bar b$, and $t \bar t$
and therefore one
can cut on large $\Delta\phi ({\rm jet}_1, {\rm jet}_2)$
to suppress the 
QCD background due to light quark flavours and $b \bar b$ as well as $t \bar t$. We 
reject isolated $e/\mu$ from background $W/Z$ leptonic decays and it is found that
the $e/\mu$ veto significantly enhances the 
GNLSP signals over the SM background. Specific cuts are given in \cite{Feldman:2009zc}.

That the LHC will turn into a gluino factory if the GNSLP model is realized is seen in Fig.(\ref{GF}).
Fig.(\ref{GF}) shows that a gluino NLSP consistent with the two sided WMAP relic density constraints can span the entire range from $ \sim $ 220 GeV
to almost a TeV over
the parameter space investigated. This analysis includes 
the radiative decay of the gluino, which can in some cases dominate the 
branching fractions of the gluino. 
As can be seen from Fig.(\ref{GF}), with just 10/fb a GNLSP can be discovered up to about 800 GeV
and thus if the gluino and neutralino masses are split over a narrow gap, this mass hierarchy
at the LHC  can be tested  into the TeV region \cite{Feldman:2009zc}.

\vspace{-.4cm}
\section{Connecting PAMELA and the LHC}
\vspace{-.2cm}
 \begin{figure*}[t]
 \centering
\includegraphics[width=6.5cm,height=5.7 cm]{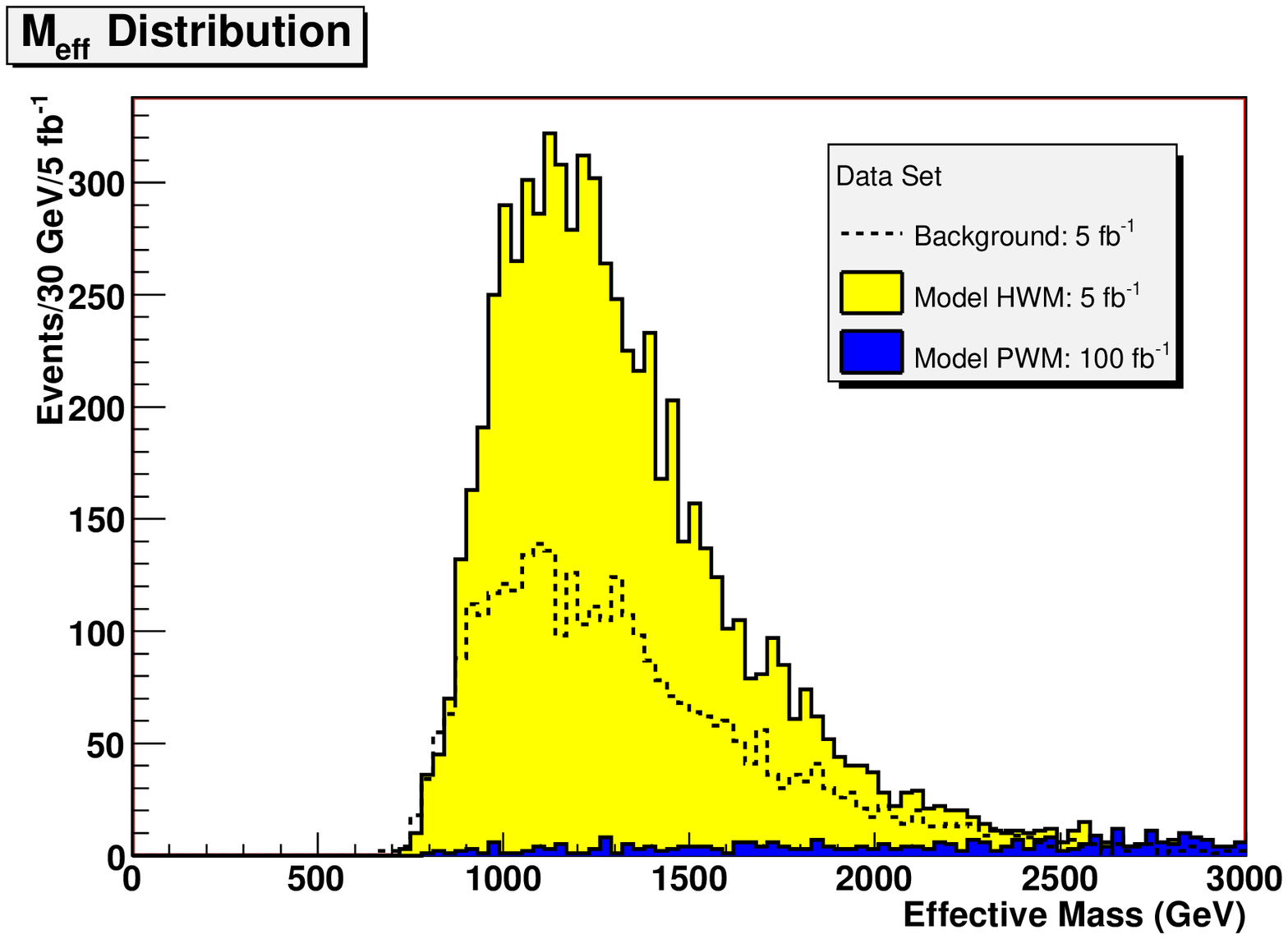}
\includegraphics[width=6.1cm,height=5.15 cm]{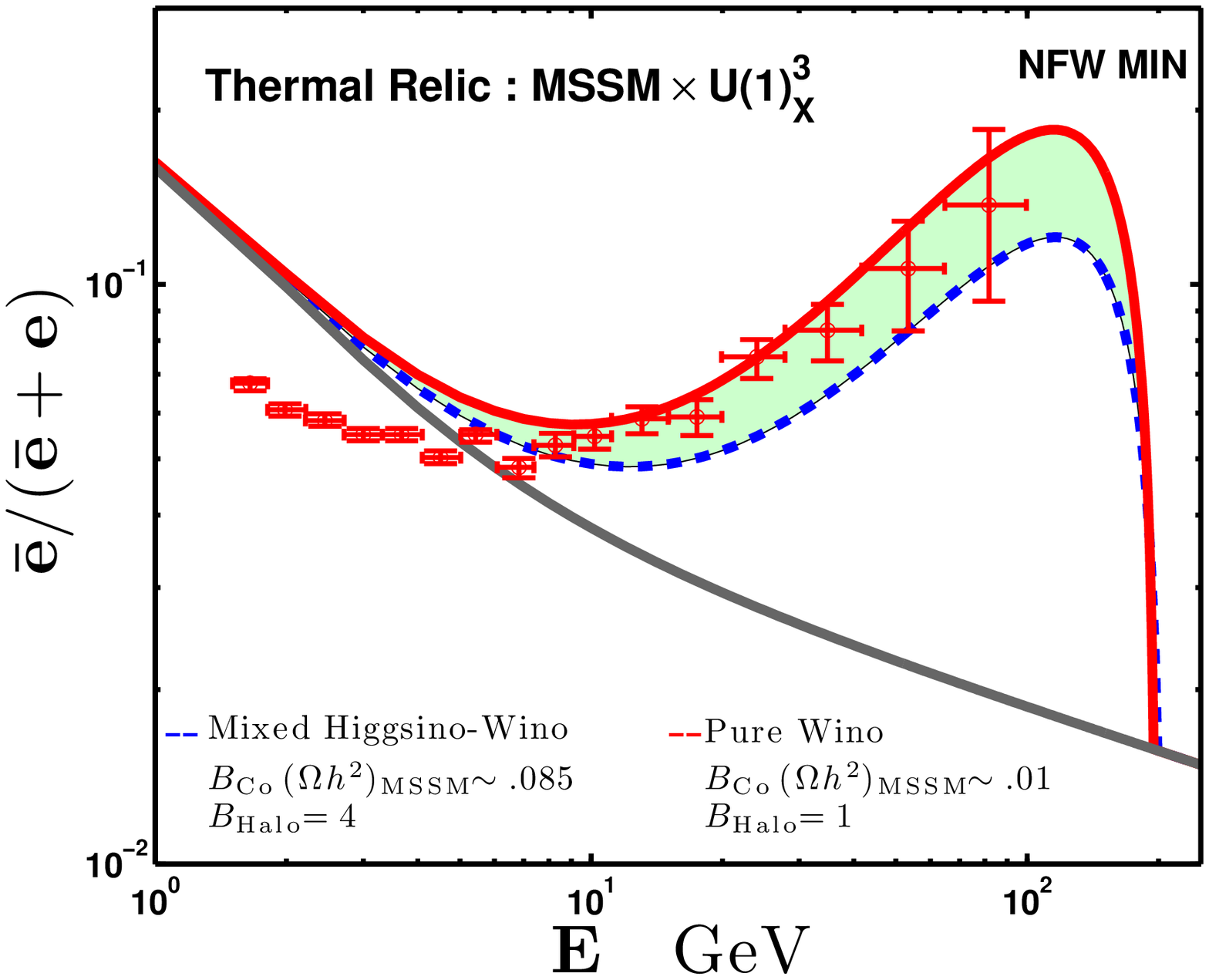}
\vspace{-.3cm}
\caption{Left: Effecive mass distribution for two models, a pure wino model (PWM) (blue/dark) which is on the floor even with $L =$100/fb in this model, and
a mixed Higgsino-Wino model (HWM) (yellow/light) which produces a strong signal with  $L =$ 5/fb. For cuts see \cite{Feldman:2009wv}. Right: Both models can fit the PAMELA positron
data with small clump factors $\lesssim  5$, while the HWM can give rise to satisfaction of the WMAP data. Both models
 and are capable of satisfying the $\bar p$ constraint for a MIN diffusion model \cite{Feldman:2009wv}.
}\label{posit}
\end{figure*}

The connection  between colliders and dark matter has become increasingly relevant
due to new data released by the PAMELA collaboration  indicating an excess in 
positron flux in the halo. In SUSY models \footnote{Discussed here are high scale models which lead to REWSB  as in the previous
sections.} with MSSM spectra 
 annihilations of $\na \na \to WW,ZZ$ are dominant possible sources of positrons in the energy range of interest.
The $WW$ production can lead to a sufficiently large  cross sections in the halo to account for the PAMELA anomaly
(for recent related work see  \cite{GKPPW,LR,Hooper:2008kv,Hisano,Feldman:2009wv}).
For a pure wino the
halo  cross section for  the $WW$ mode for $m_{\na} \sim (180 - 200) ~\rm GeV$ can be as large as $\sim 10^{-24} cm^3/s$, yet the
 mass splitting between  $m_{\cha}$ and $m_{\na}$ is squeezed to be order the pion mass when  the spectrum
is MSSM like.  However, the PAMELA data can be fit when the LSP has a non-negligible Higgsino component, and in such a case, the
mass gap between $m_{\cha}$ and $m_{\na}$ opens up \cite{Feldman:2009wv}.
A mass splitting of order 10 GeV
maintains a large enough wino component to produce a sufficiently large halo cross section, 
 and leads to a significant change in the wave function of the LSP. This effect comes about
from non universalities in gaugino masses, and can occur if there is, for example, a relatively large 
 reduction of  both the $SU(2)$ and $SU(3)$ gaugino masses  at the GUT scale relative to the pure wino, where the pure wino
can come about from a reduced $SU(2)$ gaugino mass. For this Higgsino-wino model (HWM) 
the LSP  also has a relatively large bino component (we still call it HWM to distinguish it from the pure wino)
and most importantly there is a compression of 
the colored sparticle masses opening up detectable signals at the LHC
while the PAMELA data is also fit \cite{Feldman:2009wv}
(see Fig.(\ref{posit}) for one such comparison). The Higgsino content also leads to detectable spin independent cross
section that can be observed with the CDMS and XENON-10 experiments \cite{XENONCDMS}. The mass splitting also leads to a LSP which 
suffers less from the coannihilations with the chargino. The correct relic density prediction can be obtained 
from the presence of extra $U(1)_X$ gauge factors and thus the presence of extra Majorana matter
arising from the hidden sector. 
The extra Majorana matter is naturally extra weakly interacting due to electroweak constraints.
However it supplies extra degrees of freedom to the relic abundance of the LSP through coannihilations and therefore  
{\em enhances} the relic density of the LSP neutralino. This is maximized for the HWM over the pure wino 
due to the lifting of the mass degeneracy for the HWM. This general concept of
boosting the relic density of a thermal neutralino through coannihilations 
appeared earlier in \cite{Feldman:2006wd}  as well as in \cite{Profumo:2006bx} in different contexts. 
If indeed the PAMELA data is due to neutralino annihilations in the halo,
with the contributing annihilations controlled by MSSM interactions,
the high energy $e^{+} + e^{-}$ flux reported by FERMI  must be dominantly a consequence of other astrophysical sources such
as pulsars (see e.g. \cite{Barger:2009yt,LR}).   

On the other hand, dark matter annihilations which have a significant
direct production of positrons through a pole \cite{Feldman:2008xs} 
can simultaneously explain the PAMELA data as well as the WMAP data. 
For models which directly produce leptons,
the high energy flux can  be fit within the confines of the current experimental data
(see \cite{Feldman:2008xs} and \cite{Grasso:2009ma} ).
For the case of a Breit-Wigner enhancement in the galactic halo \cite{Feldman:2008xs}
the fits to both the fluxes and WMAP data are made possible due to the spreading of $\langle \sigma v \rangle$ when integrating $\langle \sigma v \rangle$ up to the freeze out temperature in the vicinity of the pole. This may hint at a very narrow $Z'$ which couples with hypercharge enhanced couplings to 
the SM fermions, and couples vectorially through a Stueckelberg field 
to a dark dirac fermion\cite{KN}.

\vspace{-.29cm}
\section{Conclusions}
\vspace{-.29cm}
 Low lying SUSY spectra, which sit in specific mass hierarchies, reveal that SUGRA and D-brane models have early discovery prospects
at the LHC. 
These mass hierarchical patterns can be resolved by combining dark matter signatures
 and LHC signatures of new physics.
Emphasized also was the possibility of 
 a light chargino and a gluino that can surface with early
runs at the LHC.  Included in the discussion was the recently explored
class of GUT models that yield a gluino as the NLSP (GNLSP) which satisfy
WMAP constraints via gluino coannhilations and also give rise to signals
at the LHC which are overwhelmingly dominated by gluino production.
Several models reviewed have light SUSY Higgses 
and mass hierarchies involving these light Higgses
 are expected to be tested further in the next round of experiments.

Also discussed was the PAMELA positron excess which can be explained by SUSY dark matter
in high scale models. Such a prediction depends importantly on the
eigen decomposition of the LSP. A splitting of the chargino and LSP mass
plays a central role in  the observational prospects of these models.
It is found that high scale models can predict a LSP with a non-negligible Higgsino component that can fit PAMELA
and produce easily detectable LHC signatures, while a pure wino will be rather difficult to observe should the SUSY scale be order a few TeV. 
The $\bar p$ constraints on these models can be 
satisfied for a MIN diffusion model. WMAP constraints  can be accommodated in the presence of extra $U(1)_X$ factors
which lead to extra Majorana degrees of freedom of spectator states with suppressed interactions that
enhance the relic density of a mixed Higgsino-wino LSP via coannihilations.  
 We also briefly mentioned earlier  
work which can fit both the low and high energy flux data, as well as the WMAP data, via the presence of a pole\footnote{The Breit-Wigner enhancement with fits to the PAMELA data first appeared  in Ref. 1 of \cite{Feldman:2008xs} where WMAP data was
also fit (see also the last 4 Refs. of \cite{KN} \& \cite{Feldman:2007nf}). }. This comes about via
a very narrow $Z'$ resonance. A discovery of this type of $Z'$ at the Tevatron and/or LHC
in the dilepton channel may point us to a dark dirac fermion  with mass that is slightly larger
than  $M_{Z'}/2$ and thus lend further support for this distinct possibility of  leptonic annihilations in the halo \cite{KN,Feldman:2008xs}. 

{\em Acknowledgements:}
{This research received funding from
 NSF grant PHY-0757959 as well as from the Office of the Vice-Provost of Northeastern University,
and support from  the Michigan Center for Theoretical Physics (MCTP) and DOE grant DE-FG02-95ER40899.}

\end{document}